\documentclass[
reprint, 
superscriptaddress,
 amsmath,amssymb,
prl,
]{revtex4-1}

\usepackage{graphicx}
\usepackage{dcolumn}
\usepackage{bm}
\usepackage{todonotes}

\usepackage{color}

\newcommand{\pdiff}[2]{\frac{\partial #1}{\partial #2}}

\newcommand{\new}{\nonumber\\}
\newcommand{\abs}[1]{\left|#1\right|}

\newcommand{\br}{\bm{r}}

\usepackage{amsmath}	
\begin{document}

\preprint{APS/123-Qed} \title{ Jamming below upper critical dimension}


\author{Harukuni Ikeda}
 \email{hikeda@g.ecc.u-tokyo.ac.jp}
\affiliation{
Graduate School of Arts and Sciences, The University of Tokyo 153-8902, Japan
}


\date{\today}
	     
\begin{abstract}  
Extensive numerical simulations in the past decades proved that the
critical exponents of the jamming of frictionless spherical particles
are the same in two and three dimensions. This implies that the upper
critical dimension is $d_u=2$ or lower. In this work, we study the
jamming transition below the upper critical dimension. We investigate a
quasi-one-dimensional system: disks confined in a narrow channel. We
show that the system is isostatic at the jamming transition point as in
the case of standard jamming transition of the bulk systems in two and
three dimensions. Nevertheless, the scaling of the excess contact number
shows the linear scaling. Furthermore, the gap distribution remains
finite even at the jamming transition point. These results are
qualitatively different from those of the bulk systems in two and three
dimensions.
\end{abstract}


\maketitle

\paragraph*{Introduction. --}

When compressed, particles interacting with finite ranged potential
undergo the jamming transition at the critical packing fraction
$\varphi=\varphi_J$ at which particles start to touch, and the system
acquires rigidity without showing apparent structural
changes~\cite{liu2010jamming}. One of the most popular models of the
jamming transition is a system consisting of frictionless spherical
particles~\cite{ohern2003}. The nature of the jamming transition of the
model is now well understood due to experimental and numerical
investigations in the past decades~\cite{liu2010jamming}. A few
remarkable properties are the following: (i) the system is nearly
isostatic at $\varphi_J$; namely, the number of constraints is just one
greater than the number of degrees of
freedom~\cite{bernal1960,goodrich2012finite}, (ii) the excess contact
number $\delta z$ from the isostatic value exhibits the power-law
scaling $\delta z \sim \delta\varphi^{a}$ where
$\delta\varphi=\varphi-\varphi_J$ denotes the excess packing
fraction~\cite{ohern2003}, (iii) the distribution of the gap between
particles $g(h)$ exhibits the power-law divergence $g(h)\sim
h^{-\gamma}$ at $\varphi_J$~\cite{PhysRevE.71.011105}, and (iv) the
critical exponents, $a=1/2$ and $\gamma=0.41$, do not depend on the
spatial dimensions $d$ for $d\geq
2$~\cite{ohern2003,charbonneau2014fractal}.

Interestingly, the values of $a$ and $\gamma$ agree with the results of
the mean-field theories, such as the replica
method~\cite{charbonneau2014fractal,franz2016,franz2017universality},
variational argument~\cite{wyart2005compress,yan2016}, and effective
medium theory~\cite{degiuli2014effects}. This implies that the upper
critical dimension $d_u$, above which the mean-field theory provides
correct results, is $d_u\leq 2$. An Imry-Ma-type
argument~\cite{wyart2005rigidity} and recent finite-size scaling
analysis~\cite{hexner2019} also suggest $d_u\leq 2$.

\begin{figure}[t]
 \includegraphics[width=8cm]{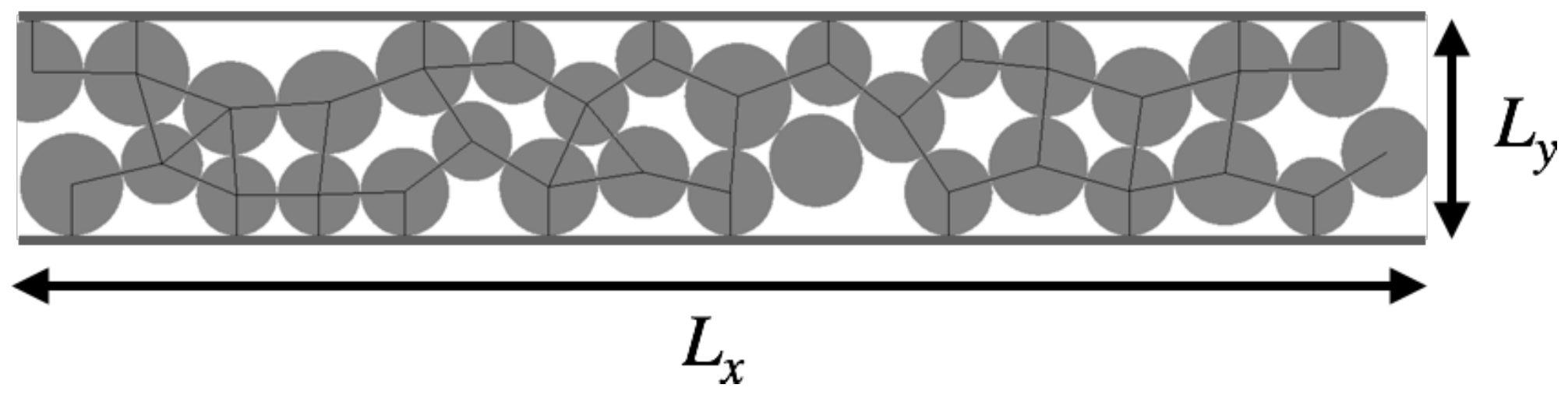} \caption{A configuration at
 $\varphi_J$ for $N=32$ and $L_y=2\sigma_{\rm max}$. Gray circles represent
 particles, and the solid lines denote the contacts.}
 \label{081731_14Feb20}
\end{figure}

A natural question is then what will happen below the upper critical
dimension. To answer this question, we here investigate the jamming
transition for $d<2$. However, the jammed configuration of a true $d=1$
system is trivial: for $\varphi\geq \varphi_J$, the number of contacts
per particle is just $z=2$, unless next nearest neighbor particles begin
to interact at very high $\varphi$. To obtain non-trivial results, we
consider a quasi-one-dimensional system as shown in
Fig.~\ref{081731_14Feb20}, where particles are confined between the
walls at $y=0$ and $y=L_y$. In the thermodynamic limit with fixed $L_y$,
the model can be considered as a one-dimensional system, but the jammed
configuration is still far from trivial.

In the previous works, quasi-one-dimensional systems have been studied
to elucidate the effect of confinement on the jamming
transition~\cite{landry2003confined,desmond2009}. These studies uncover
how the confinement changes the transition point
$\varphi_J$~\cite{desmond2009} and the distribution of the stress near
the walls~\cite{landry2003confined}. However, the
investigation of the critical properties is limited for the systems with
very small $L_y$ where the jammed configuration is similar to that of
the true $d=1$ system: each particle contact with at most two particles,
and therefore one can not discuss the scaling of $\delta
z$~\cite{ashwin2009,ashwin2013,godfrey2014}. To our knowledge, the
scaling of $\delta z$ for an intermediate value of $L_y$ has not been studied
before.

In this work, by means of extensive numerical simulations, we show that
the system is always isostatic at the jamming transition point for all
values of $L_y$, as in the case of the jamming in $d\geq 2$.
Nevertheless, the critical behavior of the jamming of the
quasi-one-dimensional system is dramatically different from the jamming
transition in $d\geq 2$.  We find that the excess contact number $\delta
z$, and the excess constraints $\delta c$, which plays a similar role as
$\delta z$, exhibit the linear scaling $\delta z\sim \delta c \sim
\delta\varphi$. Furthermore, we find that $g(h)$ remains finite even at
$\varphi_J$. These results prove that the jamming transition of the
quasi-one-dimensional system indeed shows the distinct scaling behaviors
from those in $d\geq 2$.

\paragraph*{Model. --}
Here we describe the details of our model. We consider two dimensional
disks in a $L_x\times L_y$ box. For the $y$-direction, particles are
confined between the walls at $y=0$ and $y=L_y$. For the $x$-direction,
we impose the periodic boundary condition. The interaction
potential of the model is given by
\begin{align}
& V_N = \sum_{i<j}^{1,N}v(h_{ij}) + \sum_{i=1}^N v(h_i^{\rm b})
 + \sum_{i=1}^N v(h_i^{\rm t}),\new
 &h_{ij}=\abs{\br_i-\br_j}-\frac{\sigma_i+\sigma_j}{2},\new
&h_i^{\rm b}=y_i-\frac{\sigma_i}{2},\ h_i^{\rm t}=L_y-y_i-\frac{\sigma_i}{2},\new
&v(h)=k \frac{h^2}{2}\theta(-h),\label{084239_24May20}
\end{align}
where $\br_i=\{x_i,y_i\}$ and $\sigma_i$ respectively denote the
position and diameter of particle $i$, $h_{ij}$ denotes the gap function
between particles $i$ and $j$, and $h_i^{\rm b}$ and $h_i^{\rm t}$ respectively
denote the gap functions between particle $i$ and bottom and top walls.
To avoid crystallization, we consider polydisperse particles with
uniform distribution $\sigma_i\in [\sigma_{\rm min},\sigma_{\rm
max}]$. Here after we set, $k=1$, $\sigma_{\rm min}=1$, and $\sigma_{\rm
max}=1.4$.

\paragraph*{Numerics. --}
We perform numerical simulations for $N=1024$ disks. We find $\varphi_J$
by combining slow compression and decompression as
follows~\cite{ohern2003}. We first generate a random initial
configuration at a small packing fraction $\varphi=0.1$ between the
walls at $y=0$ and $y=L_y$. Then, we slowly compress the system by
performing an affine transformation along the $x$-direction. For each
compression step, we increase the packing fraction with a small
increment $\delta \varphi=10^{-3}$, and successively minimize the energy
with the FIRE algorithm~\cite{fire2006} until the squared force acting
on each particle becomes smaller than $10^{-25}$. After arriving at a
jammed configuration with $V_N/N > 10^{-16}$, we change the sign and
amplitude of the increment as $\delta\varphi\to -\delta\varphi/2$. Then,
we decompress the system until we obtain an unjammed configuration with
$V_N/N < 10^{-16}$. We repeat this process by changing the sign and
amplitude of the increment as $\delta\varphi\to -\delta\varphi/2$ every
time the system crosses the jamming transition point. We terminate the
simulation when $V_N/N\in (10^{-16},2\times 10^{-16})$. We define
$\varphi_J$ as a packing fraction at the end of the above algorithm.

After obtained a configuration at $\varphi_J$, we re-compress the system
to obtain configurations above $\varphi_J$. As reported in
Ref.~\cite{vanderwerf2020}, some fraction of samples become unstable
during the compression (compression unjamming). We neglect these
samples. We remove the rattlers that have less than three contacts
before calculating physical quantities. Hereafter, we refer the number of the
non-rattler particles as $N_{\rm nr}$. To improve the statistics, we average
over $50$ independent samples.


\paragraph*{$\varphi_J$ and $z_J$. --}
\begin{figure}[t]
 \includegraphics[width=9cm]{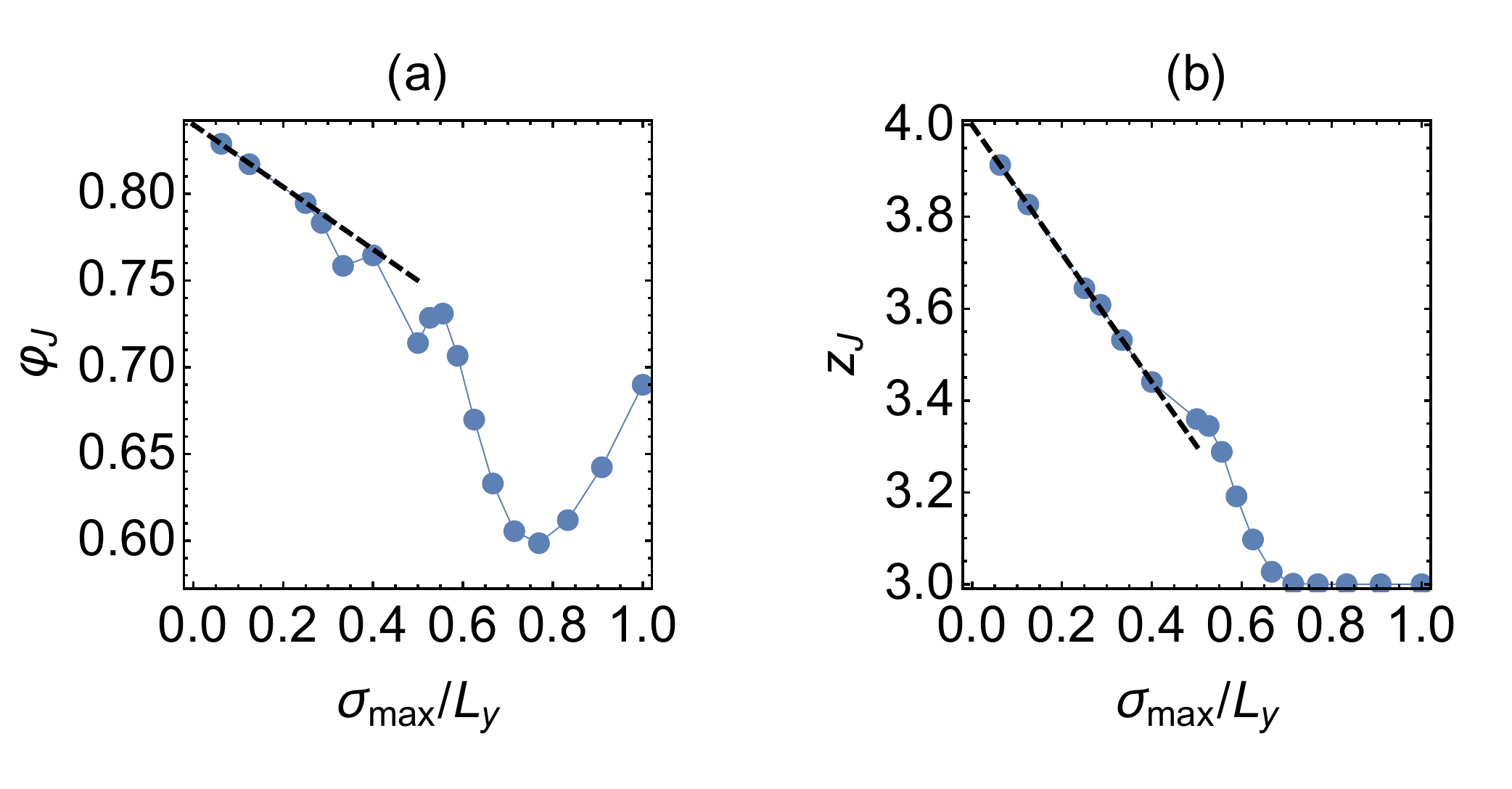} \caption{$L_y$ dependence of (a)
 the jamming transition point $\varphi_J$ and (b) the contact number per
 particle at the jamming transition point $z_J$. Markers denote
 numerical results, and solid lines denote the guide to the eye. The
 dashed lines denote the linear fits $\varphi_J=0.84-0.28\sigma_{\rm
 max}/L_y$ and $z_J=4-1.4\sigma_{\rm max}/L_y$.}
				       \label{113610_14Feb20}
\end{figure}
First, we discuss the $L_y$ dependence of the jamming transition point
$\varphi_J$ and the contact number par particle at that point $z_J$. In
Fig.~\ref{113610_14Feb20} (a), we show $\varphi_J$ as a function of
$\sigma_{\rm max}/L_y$. For intermediate values of $\sigma_{\rm
max}/L_y$, $\varphi_J$ shows a non-monotonic behavior. A similar
non-monotonic behavior has been reported in a previous numerical
simulation for a binary mixture~\cite{desmond2009}. In the limit
$\sigma_{\rm max}/L_y \to 0$, $\varphi_J$ converges to its bulk value
$\varphi_J^{\rm bulk}=0.84$ as $\varphi_J^{\rm bulk}-\varphi_J\propto
1/L_y$, see the dashed line in Fig.~\ref{113610_14Feb20} (a). The same
scaling has been observed in the previous simulation for the binary
mixture~\cite{desmond2009}. The scaling implies the growing length
scale
 $\xi \sim (\varphi_J^{\rm bulk}-\varphi)^{-\nu}$
with $\nu=1$. It is worth mentioning that this is the same exponent
observed by a correction to scaling analysis~\cite{daniel2011} and
also our replica calculation for a confined system~\cite{ikeda2015one}.

In Fig.~\ref{113610_14Feb20} (b), we show $z_J$ as a function of
$\sigma_{\rm max}/L_y$. It is well known that $z_J = z_J^{\rm bulk} = 4$
for bulk two dimensional disks~\cite{ohern2003}. In the $L_y\to\infty$
limit, $z_J$ converges to the bulk value as $z_J^{\rm bulk}-z_J \sim
1/L_y$, see the dashed line in Fig.~\ref{113610_14Feb20} (b).

\paragraph*{Isostaticity. --} 
Next we discuss the isostaticity of our model at $\varphi_J$. The
number of degrees of freedom of the non-rattler particles is $N_f =
2N_{\rm nr}-1$ where $N_{\rm nr}$ denotes the number of non-rattler
particles, and we neglect the global translation along the $x$-axis. The
number of constrains is
\begin{align}
N_c =
 \frac{N_{\rm nr}z-N_{\rm w}}{2} + N_{\rm w}
=  \frac{N_{\rm nr}z}{2} + \frac{N_{\rm w}}{2},
\end{align}
where $z$ denotes the number of contacts per particle, $N_{\rm w}$
denotes the number of contacts between particles and walls, and $(N_{\rm
nr}z-N_{\rm w})/2$ accounts for the number of contacts between
particles.
\begin{figure}[t]
 \includegraphics[width=6cm]{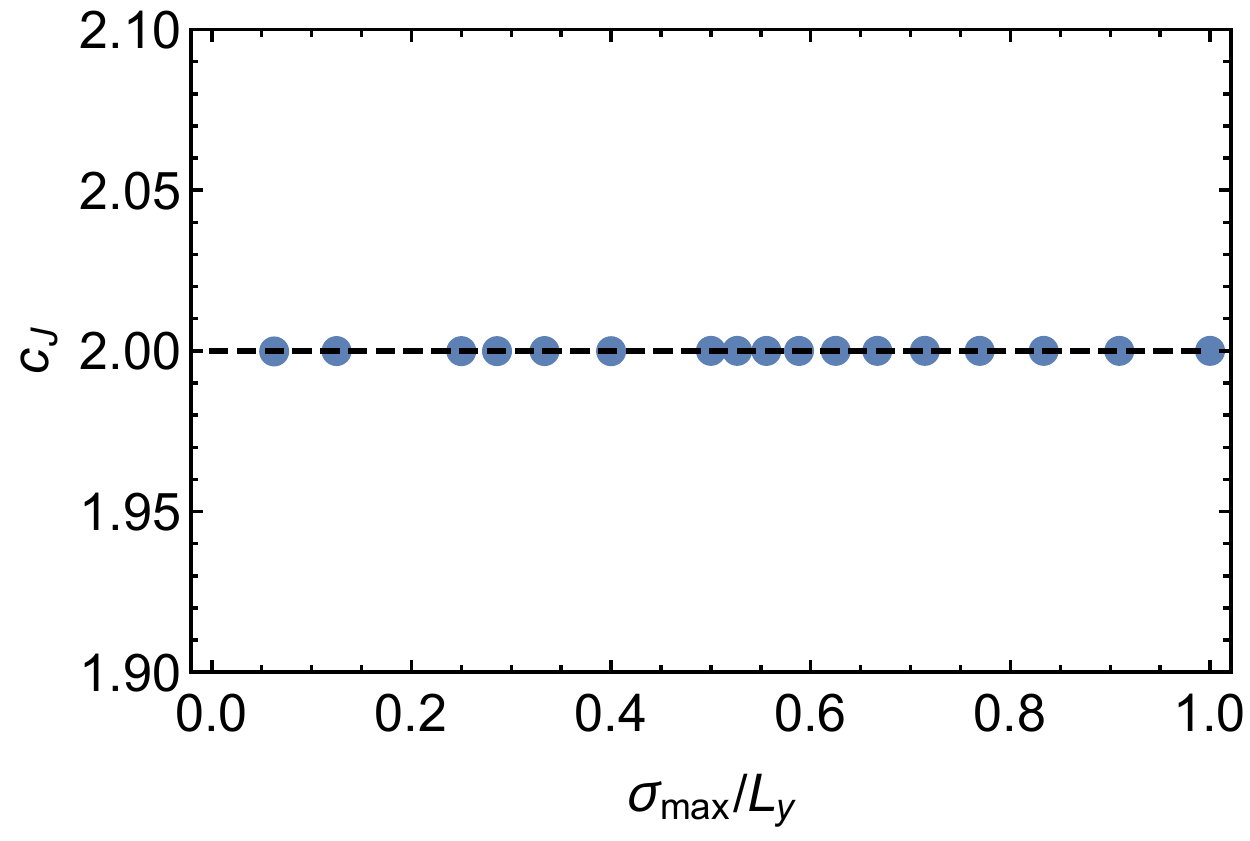} \caption{$L_y$ dependence of the
 number of constraints per particle at the jamming transition point
 $c_J$. Markers denote the numerical results, and the dashed line
 denotes the isostatic number $c_{\rm iso}=2$.}  \label{130327_14Feb20}
\end{figure}
To discuss the isostaticity, we observe the number of constraints per
particle $c= N_c/N_{\rm nr}$. When the system is isostatic $N_c=N_f$, we
get $c=c_{\rm iso}=2$ in the thermodynamic limit.  In
Fig.~\ref{130327_14Feb20}, we show our numerical result of $c$ at
$\varphi_J$ as a function of $\sigma_{\rm max}/L_y$. This plot proves
that the system is always isostatic, irrespective of the value of $L_y$.

Now we shall discuss the behavior above $\varphi_J$. As mentioned in the
introduction, we will investigate the model mainly for $L_y >
2\sigma_{\rm min}$ so that some fraction of disks can pass through, and
thus the contact network undergoes a non-trivial rearrangement on the
change of $\varphi$.

\paragraph*{Energy and pressure. --}

For $\varphi>\varphi_J$, the particles overlap each other. As a
consequence, the energy $V_N$ and pressure $p$ have finite values. Since
we only consider the compression along the $x$-axis, we define the
pressure as
\begin{align}
 p = -\frac{1}{V}\left.\pdiff{V_N(\{x_i'\})}{\varepsilon}\right|_{\varepsilon=0}
 = -\frac{1}{V}\sum_{i<j}v'(h_{ij})\frac{(x_i-x_j)^2}{\abs{\br_i-\br_j}},
\end{align}
where $V=L_x L_y$, and $x_i' = x_i(1+\varepsilon)$ denotes the affine
transformation along the $x$-axis.
\begin{figure}[t]
 \includegraphics[width=9cm]{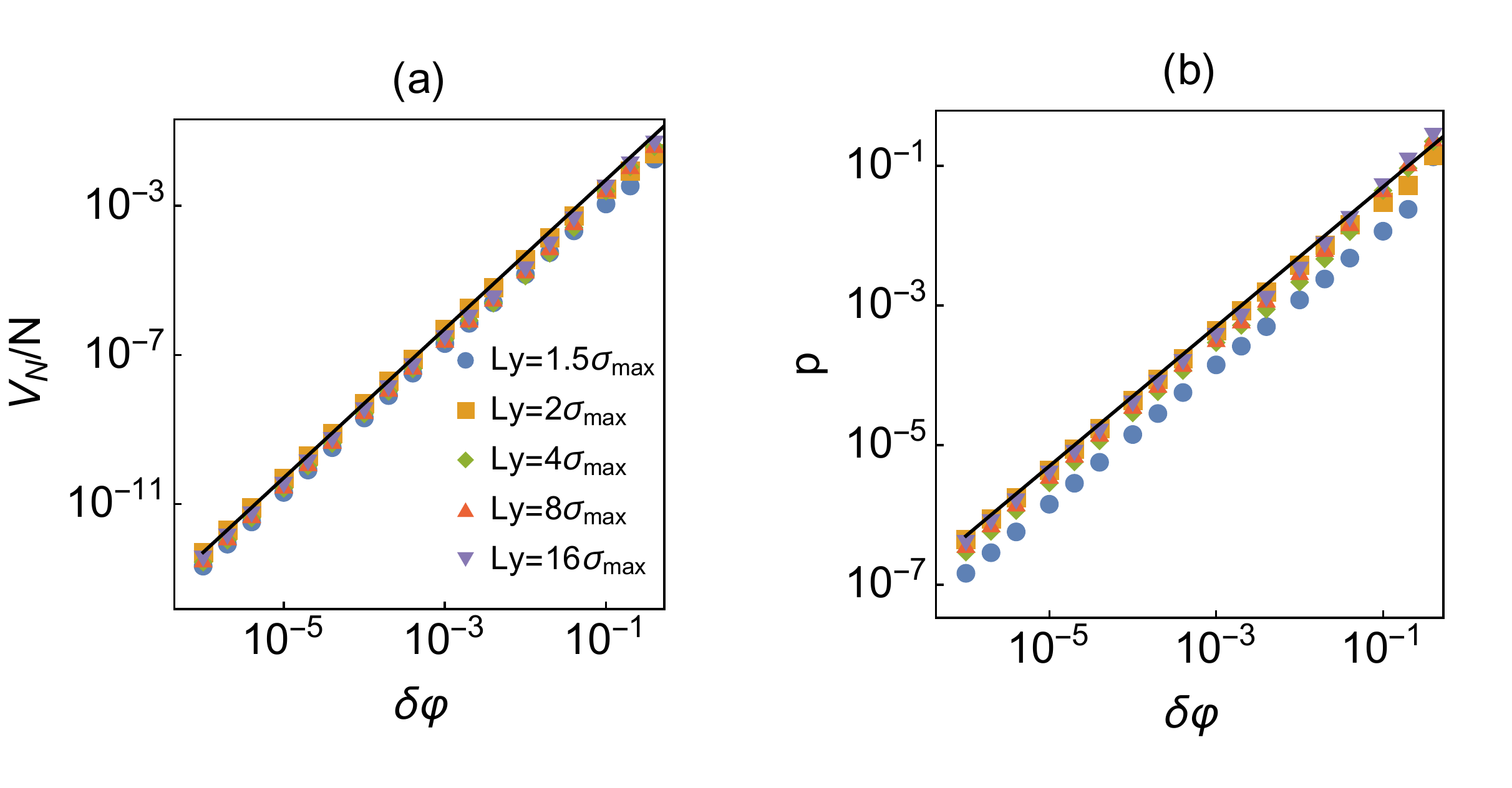} \caption{(a) $\delta\varphi$
 dependence of the energy per particle $V_N/N$.  Maker denote numerical
 results, and the solid line denotes $\delta\varphi^2$.  (b)
 $\delta\varphi$ dependence of the pressure $p$. Maker denote numerical
 results, and the solid line denotes $\delta\varphi$.}
					 \label{063704_16Feb20}
\end{figure}
In Fig.~\ref{063704_16Feb20}, we show the $\delta\varphi$ dependence of
$V_N/N$ and $p$.  We find the scalings $V_N/N \sim
\delta\varphi^2$ and $p\sim \delta\varphi$. The same scalings were
observed for the bulk systems in $d=2$ and $d=3$~\cite{ohern2003}.

\paragraph*{Number of constraints and contacts. --}
Next we observe the density dependence of the number of constraints.
For this purpose, we introduce the excess constraints as
\begin{align}
 \delta c = \frac{N_c-(N_f+1)}{N_{\rm nr}}.
\end{align}
where $N_f+1$ denotes the minimal number of constraints to stabilize a
system consisting of frictionless spherical
particles~\cite{wyart2005rigidity,goodrich2012finite}. For the bulk
limit $L_{y}\sim \sigma_{\rm max}\sqrt{N}$, $\delta c$ can be identified
with the excess contact number $\delta z$. In this case, the extensive
finite size scaling analysis proved the following scaling
form~\cite{goodrich2012finite}:
\begin{align}
 \delta c = 
 {N}^{-1} \mathcal{C}\left({N}^2\delta\varphi\right),\label{122232_1Mar20}
\end{align}
where the scaling function $\mathcal{C}(x)$ behaves as
\begin{align}
 \mathcal{C}(x) \sim
 \begin{cases}
  x^{1/2} & x\gg 1\\
  x & x\ll 1.
 \end{cases}\label{140435_3Mar20}
\end{align}
This implies that the square root behavior $\delta c \sim
\delta\varphi^{1/2}$ is truncated at $\delta\varphi\sim N^{-2}$ for a
finite $N$ system. For ${\delta\varphi\ll N^{-2}}$, one observes a
linear scaling behavior ${\delta c \sim
N\delta\varphi}$~\cite{daniel2011}.

\begin{figure}[t]
 \includegraphics[width=9cm]{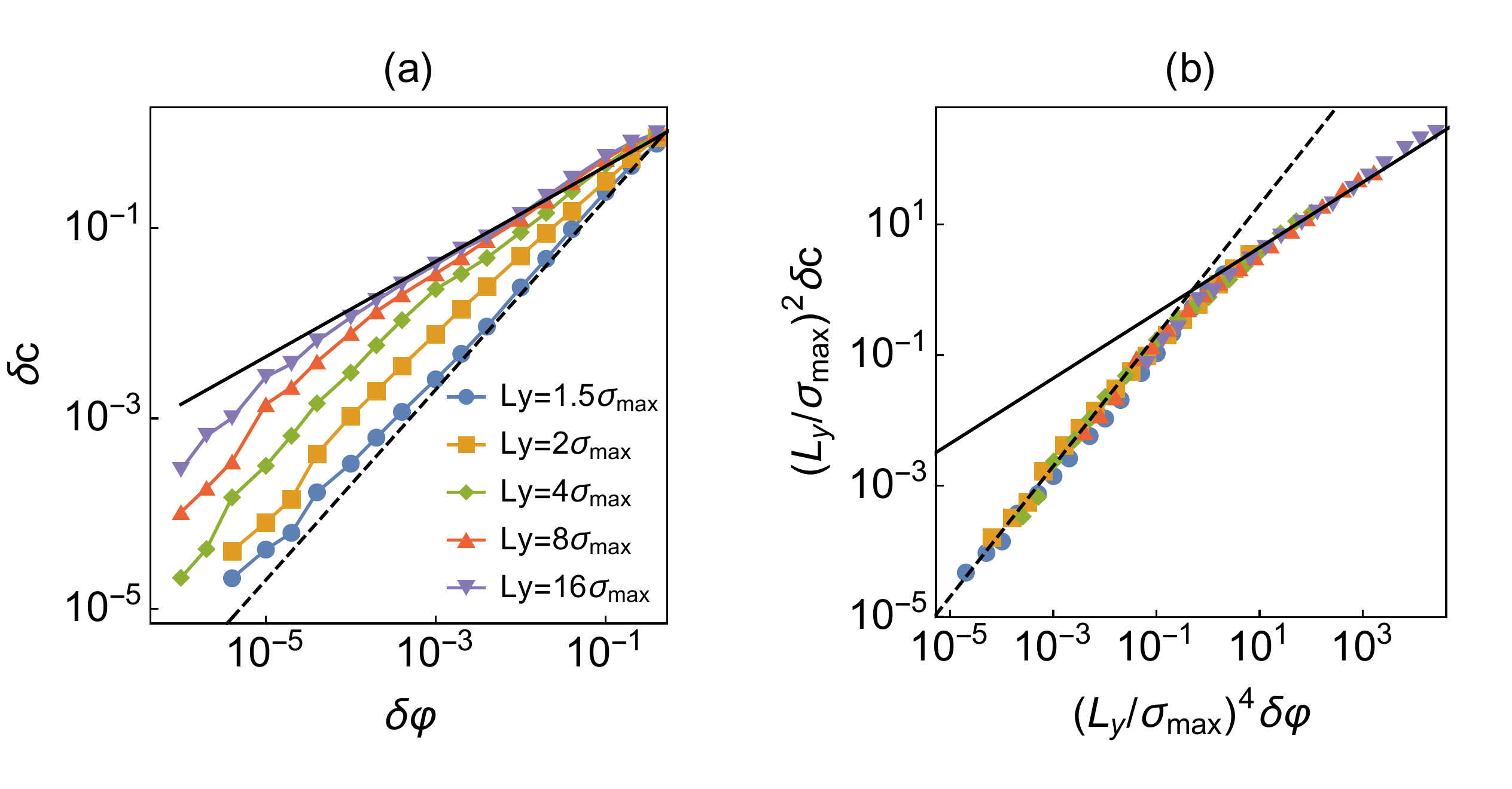} \caption{(a) $\delta c$ as a
 function of $\delta\varphi$.  Markers denote numerical results. The
 solid and dashed lines denote $\delta c\sim \delta\varphi^{1/2}$ and
 $\delta c\sim \delta\varphi$, respectively. (b) Scaling plot for the
 same data.}  \label{140943_14Feb20}
\end{figure}
To investigate how the behavior changes for $L_y\ll
\sigma_{\rm max}\sqrt{N}$, in Fig.~\ref{140943_14Feb20} (a), we show the
$\delta\varphi$ dependence of $\delta c$ for several $L_y$. For large
$L_y$ and intermediate $\delta\varphi$, we observe the square root
scaling $\delta c\sim \delta\varphi^{1/2}$. On the contrary, for small
$L_y$ and $\delta\varphi$, $\delta c$ shows the linear behavior $\delta
c \sim \delta\varphi$. To discuss the scaling behavior more closely, we
assume the following scaling form: 
\begin{align}
 \delta c =
 l_y^\alpha \mathcal{C}'\left(l_y^\beta\delta\varphi\right),\label{114128_23Feb20}
\end{align}
where $l_y=L_y/\sigma_{\rm max}$, and $\mathcal{C}'(x)$ shows the same
scaling behavior as $\mathcal{C}(x)$, Eq.~(\ref{140435_3Mar20}). When
$l_y\sim \sqrt{N}$, the scaling should converge to that of the bulk
$d=2$ system, Eq.~(\ref{122232_1Mar20}). This requires $\alpha=-2$ and
$\beta=4$. In Fig.~\ref{140943_14Feb20}, we test this prediction. A
good scaling collapse verifies the scaling function
Eq.~(\ref{114128_23Feb20}).

Note that for a bulk system in $d\geq 2$, the system exhibits the linear
scaling only for $\delta\varphi\ll N^{-2}$: the linear regime vanishes
in the thermodynamic limit.  Contrary, Eq.~(\ref{114128_23Feb20})
implies that the linear scaling regime persists even in the
thermodynamic limit for the quasi-one-dimensional system as long as
$L_y$ is finite. Therefore, the quasi-one-dimensional system indeed has
a distinct critical exponent from that of the bulk systems in $d\geq 2$.

\begin{figure}[t]
 \includegraphics[width=9cm]{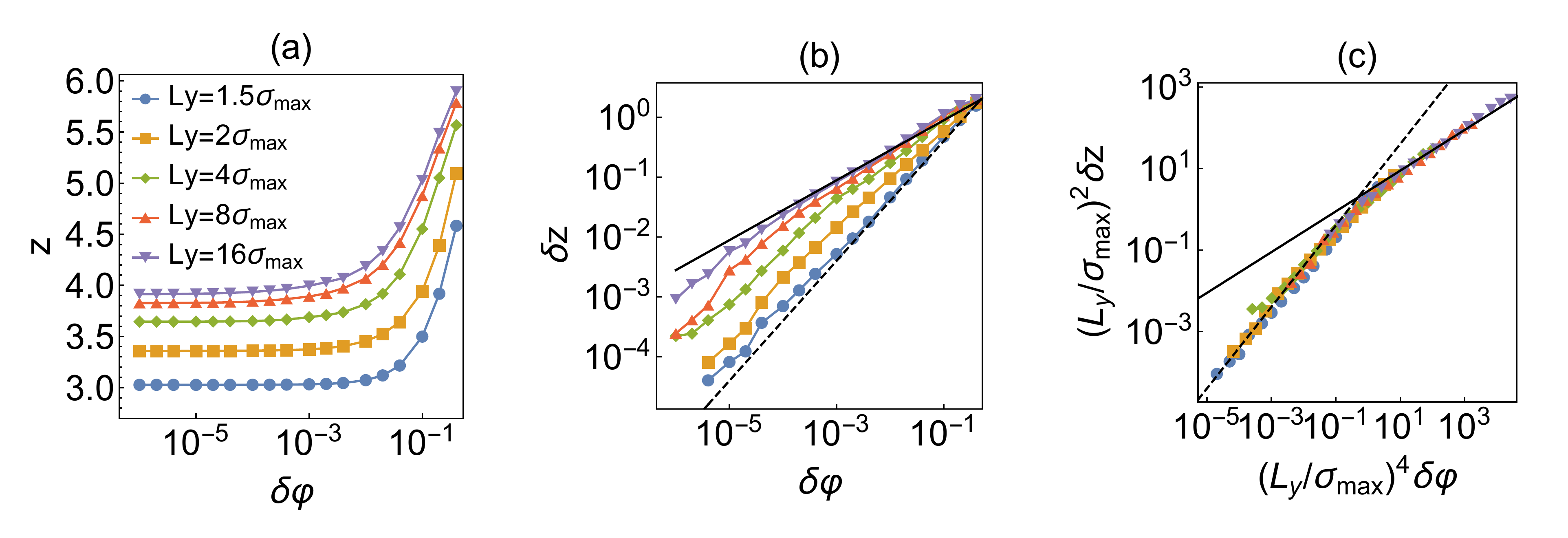} \caption{
(a) $z$ as a
 function of $\delta\varphi$. Markers denote numerical results. 
(b) $\delta z=z-z_J$ as a
 function of $\delta\varphi$.
The solid and dashed lines denote $\delta z\sim \delta\varphi^{1/2}$ and
 $\delta z\sim \delta\varphi$, respectively.  (c) Scaling plot for the
 same data.}
\end{figure}
In Figs.(a)--(c), we also show the behaviors of the contact number per
particle $z$, excess contacts $\delta z=z-z_J$, and its scaling
plot. The data for $\delta z$ are more noisy than $\delta c$, presumably
due to the fluctuation of $z_J$, but still we find a reasonable scaling
collapse by using the same scaling form as $\delta c$.

\paragraph*{Gap distribution. --}
Another important quantity to characterize the critical property of the
jamming transition is the gap distribution $g(h)$. For the bulk systems
in $d\geq 2$, $g(h)$ exhibits the power-law divergence at $\varphi_J$:
\begin{align}
 g(h)\sim h^{-\gamma}\label{144301_15Feb20}
\end{align}
with $\gamma=0.41$~\cite{charbonneau2014fractal}.
\begin{figure}[t]
 \includegraphics[width=9cm]{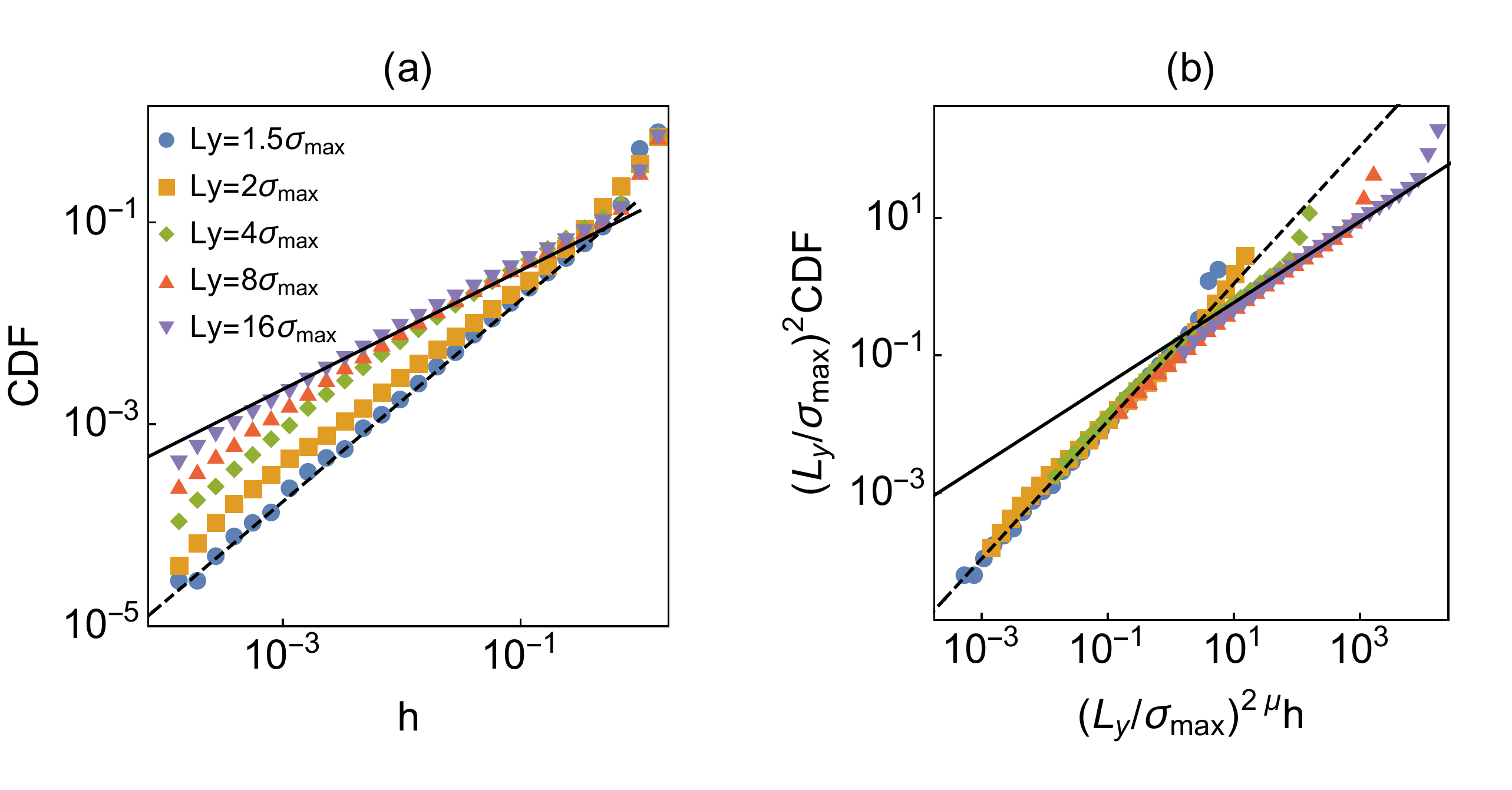} \caption{(a) CDF of the gap
 function $h$. Markers denote numerical results. The solid and dashed
 lines denote $h^{1-\gamma}$ and $h^1$, respectively. (b) Scaling plot
 for the same data.}  \label{152804_14Feb20}
\end{figure}
In order to improve the statistics, we observe the cumulative
distribution function (CDF) of the gap functions ($h_{ij}$ and $h_i^{\rm
t,b}$), instead of $g(h)$ itself.  In this case, the power-law divergence
Eq.~(\ref{144301_15Feb20}) appears as ${\rm CDF}\sim h^{1-\gamma}$. In
Fig.~\ref{152804_14Feb20} (a), we show our numerical results of CDF for
several $L_y$. We find that for small $L_y$ and $h$, ${\rm CDF}\sim h$
meaning that $g(h)$ remains finite $g(h)\sim h^0$ even at
$\varphi_J$. On the contrary, for large $L_y$, there appears the
intermediate regime where ${\rm CDF}\sim h^{1-\gamma}$, as in $d\geq 2$.
To discuss the crossover from ${\rm CDF}\sim h$ to ${\rm CDF}\sim
h^{1-\gamma}$, we assume the following scaling form:
\begin{align}
 {\rm CDF} =
 l_y^{\zeta} \mathcal{F}'\left(l_y^{\eta}h\right),\label{145121_15Feb20}
\end{align}
where the scaling function $\mathcal{F}'(x)$ behaves as
\begin{align}
 \mathcal{F}'(x) \sim
 \begin{cases}
  x^{1-\gamma} & x\gg 1\\
  x & x\ll 1.
 \end{cases}\label{140625_3Mar20}
\end{align}
When $l_y \sim \sqrt{N}$, this should converge to the scaling form for
finite $N$, ${\rm CDF}(h)= N^{-1}\mathcal{F}(N^{\mu} h)$, where
${\mu=1/(1-\gamma)}$, and $\mathcal{F}(x)$ shows the same scaling as
Eq.~(\ref{140625_3Mar20})~\cite{ikeda2019}. This requires $\zeta=-2$ and
$\eta=2\mu$. In Fig.~\ref{152804_14Feb20} (b), we check this prediction.
The excellent collapse of the data for $h\ll 1$ proves the validity of
our scaling Ansatz Eq.~(\ref{145121_15Feb20})~\footnote{Note that our
scaling prediction does not work for $h\sim 1$, where CDF does not show
the power-law behavior~\cite{PhysRevLett.109.205501}.}.

\paragraph*{Quasi-two-dimensional system. --}

\begin{figure}[t]
 \includegraphics[width=9cm]{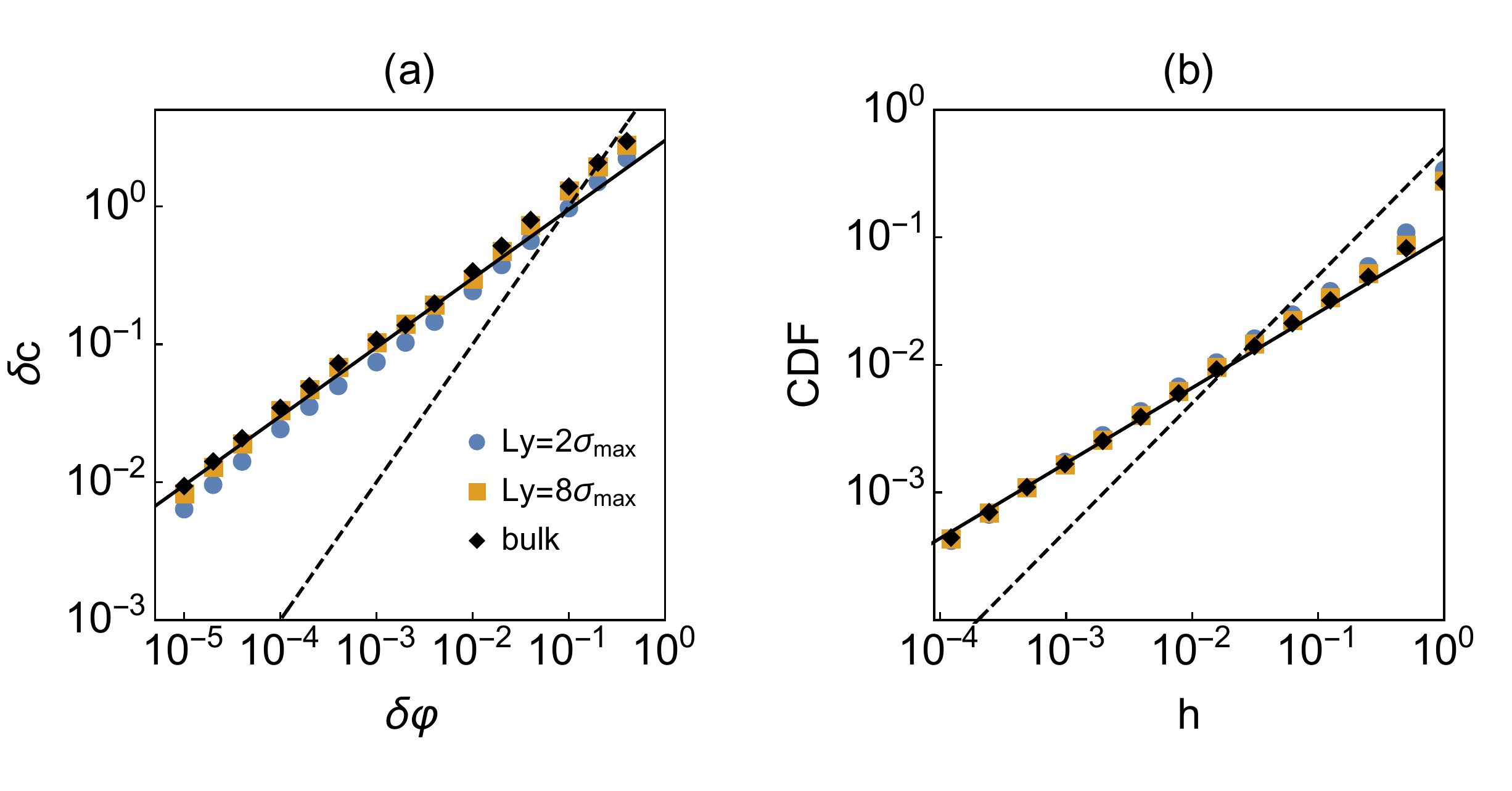} \caption{Results for the
 quasi-two-dimensional system and bulk three dimensional system.  (a)
 $\delta c$ as a function of $\delta\varphi$.  Markers denote numerical
 results. The solid and dashed lines denote $\delta c\sim
 \delta\varphi^{1/2}$ and $\delta c\sim \delta\varphi$, respectively.
(b) CDF of the gap
 function $h$. Markers denote numerical results. The solid and dashed
 lines denote ${\rm CDF}\sim h^{1-\gamma}$ and ${\rm CDF}\sim h^1$, respectively.}
				       \label{085346_24May20}
\end{figure}

One may suspect that the distinct scaling of the quasi-one-dimensional
system is due to the effect of the boundary condition, not the spatial
dimensions. To investigate this possibility, we conduct a numerical
simulation for a quasi-\textit{two}-dimensional system. We consider the
same interaction potential as Eq.~(\ref{084239_24May20}) with the same
system size $N=1024$ and polydispersity $\sigma_i\in [1.0,1.4]$, but
this time we consider spheres in a $L_x\times L_y\times L_z$ box. As
before, particles are confined between the walls at $y=0$ and $y=L_y$,
and the periodic boundary conditions are imposed along the $x$ and $z$
directions. We fix $L_y$ and change $L_x=L_z$ to control $\varphi$.  For
comparison, we also perform numerical simulations for the bulk three
dimensional system, where $L_x=L_y=L_z$ and the periodic boundary
conditions are imposed for all directions. In
Fig.~\ref{085346_24May20}, we summarize our results for $\delta c$ and
CDF of the gaps. One can see that the scaling of the
quasi-two-dimensional system is the same as that of the bulk three
dimensional system. This result implies that the different scaling of
the quasi-one-dimensional system is indeed a consequence of the fact
that one dimension is lower than the upper critical dimension.

\paragraph*{Conclusions. --}

In this work, we showed that the jamming transition in a
quasi-one-dimensional system is qualitatively different from that in
$d\geq 2$ systems: the excess constraints and contacts exhibit the
linear scaling $\delta c\sim \delta z \sim \delta\varphi$, instead of
the square root scaling $\delta z\sim\delta\varphi^{1/2}$, and the gap
distribution $g(h)$ remains finite even at $\varphi_J$, instead of the
power-law divergence $g(h)\sim h^{-\gamma}$.

Important future work is to test the robustness of our results for other
shapes of the quasi-one-dimensional geometries such as a $d$-dimensional
box with an infinite length in only one direction and fixed lengthes in
the other $d-1$ directions, and circular cylinder with a fixed radius.


\begin{acknowledgments}
\paragraph*{Acknowledgements. --}
We warmly thank M.~Ozawa, A.~Ikeda, K.~Hukushima, Y.~Nishikawa,
F.~Zamponi, P.~Urbani, and M.~Moore for discussions related to this
work. We would like in particular to thank the anonymous referee and
M.~Ozawa for suggesting the numerical simulation of the
quasi-two-dimensional system. This project has received funding from the
European Research Council (ERC) under the European Union's Horizon 2020
research and innovation program (grant agreement
n.~723955-GlassUniversality) and JSPS KAKENHI Grant Number JP20J00289.
\end{acknowledgments}


\bibliography{apssamp}
\end{document}